# Powerful evidences for supporting the claim that gamma-ray burst redshift is gravity-generated

Fu-Gao Song [†][‡]

**Abstract.** At present, it is widely believed that the phenomenon of the gamma-ray burst redshift is cosmological origin. From a theoretical point of view, this redshift has either a cosmological or a cause that is related to gravity. However, the question of whether the gamma-ray burst redshift has a cosmological origin or not should be answerable in no uncertain terms because both the spectrum characteristics and the count distribution law arising from the two distinct settings are completely different. If the redshift of GRB is generated by gravity, then the afterglow spectrum will certainly contain both the gravitational redshits (containing emission and absorption feature) and Doppler absorption redshift, and hold a definite relation between the two redshifts. In this paper, we present nine direct and decisive evidences to show that the gamma-ray burst redshift is indeed generated by gravity of neutron stars in their merging process; in which, 114 GRBs' redshifts showed that the statistical count distribution law for the two kinds redshift is the same (with errors less than 1.5%), and 74 spectral line redshifts of two GRBs showed that the relation between the two kinds redshift is completely correct (with errors less than 0.0061%).

The phenomenon of γ-ray bursts (GRB) was first observed in the late 1960's with official reports being published several years later [1,2]. The true nature and origin of this phenomenon, however, has remained a mystery owing to the great difficulty of GRB observations. The launch of the Compton Gamma-Ray Observatory (CGRO) in 1991 opened up a new era in GRB research with over 2700 bursts being recorded by the all-sky survey of the Burst and Transient Source Experiment (BATSE) on CGRO. The observations showed that the angular distribution of GRBs is generally isotropic [3–6], implying either that the γ-ray bursts are located in an extended galactic halo [7–11] or that they are cosmological in origin [12–15].

[†] College of Electronic Science and Technology, Shenzhen University, Shenzhen, 518060, China
[‡] Shenzhen Key Laboratory of Micro-nano Photonic Information Technology, Shenzhen, China

A decisive change occurred in GRB research in 1997 when the Italian-Dutch satellite Beppo-SAX was successfully launched and the redshifts for a number of GRBs were subsequently discovery [16–23] (see also Table 1 of [24]). In fact, when these redshifts were first observed, they were thought to provide irrefutable evidence for the cosmological origin of GRB because of the obvious connection between these radiation shifts and the recession of the stars that generate the radiation. Currently, it is widely accepted that GRBs are transitory physical processes of extremely high energy that take place in faraway host galaxies although it has never been confirmed. The observed redshift is therefore also called the "host redshift", and it is possible to determine the distance of the GRB from the earth by measuring the redshift. This in a nutshell is the present state of our knowledge for γ-ray bursts and the status quo of our understanding for this astronomical phenomenon! It is worth noting in particular that the prevalent belief in the cosmological origin of GRB redshifts has in fact never been strictly justified by theory.

Are GRB redshifts indeed cosmological? The answer is negative. In this paper, we exhibit a lot of evidences to support this conclusion. In particular, we present nine direct and decisive evidences to show that GRB redshifts are generated by the gravity of neutron stars and are unrelated to its distances.

It is well known that there are three kinds of redshifts in nature, namely, the cosmological redshift, the gravitational redshift and the Doppler redshift arising from local motion. If gravity produces GRB redshifts, then it will also produce Doppler redshifts that arise from the local motion caused by the gravitational field. On the other hand, if GRB redshifts are indeed cosmological, then it may arise from either the host galaxy or the background galaxy. The problem of determining the real cause of GRB redshifts may therefore seem quite difficult to tackle. However, owing to the many distinct statistical distribution features of number-redshift relation between the two types of origin, the true cause of GRB redshift can indeed identified.

Now it is well known that quasars come from external galaxies and hence have redshift. We are already familiar with the number-redshifts relation of quasars, i.e. $N_{\text{quasar}}(z) = c z^3$ for lower redshift [25], where $N_{\text{quasar}}(z)$ is the number of quasars with redshifts not exceeding $z$, $c$ a constant. If the origin of GRBs is indeed cosmological, then its distribution must be the same as quasars, but the expectation does not occur. We actually have $N_{\text{GRB}}(z) = c z$ for a large redshift (see Fig. 1 below), which means that the redshift of GRBs has nothing to do with its distance.

What are the particular features of redshift of GRBs when it results from gravity? We shall discuss this problem now, and specially describe the redshift features of a GRB event taking place on the surface of a neutron star that is formed by the merging of two neutron stars in a close binary neutron star system.

It is known that the gravitational redshift $z_G$ of spectral line for an atom resting on the surface of a neutron star with mass $M$ and radius $r_0$ is

$$z_G = (1 - a/r_0)^{-1/2} - 1 = (1 - \kappa)^{-1/2} - 1, \tag{1}$$

where, $a = 2GM/c^2$, $\kappa = a/r_0$, $G$ is the gravitational constant, and $c$ is the velocity of light. On the other hand, it is also known that the Doppler redshift $z_d$ of spectral line for an atom moving with a recession velocity $v$ is

$$z_d = [(1+\beta)/(1-\beta)]^{1/2} - 1, \quad \beta = v/c. \tag{2}$$

Now, when enormous quantities of matter are spurted out in a GRB process, only a small amount of it would eventually return to the star. The returning matter would attain great speed and then produce a large Doppler redshift in the spectral lines of the absorption spectrum. For this reason, therefore, absorption spectral lines are often observed in a GRB process in addition to the gravitational redshifts.

The Doppler redshift can be easily deduced. Recall that the rest energy of an electron with a rest-mass $m_0$ is $m_0c^2$, this energy would become $m_0c^2(1-a/r_1)^{1/2}$ when its atom is at rest in a gravitational field at a distance $r_1$ from the centre of a star of mass $M$. Let $v$ be the macroscopic speed attained by the electron after free fall to the surface of the star with radius $r_0$. Then the electron's energy at the time of impact is $m_0c^2(1-\kappa)^{1/2}/(1-\beta^2)^{1/2}$, where $m_0c^2(1-\kappa)^{1/2}$ is the electron's rest energy at the star's surface if it is at rest. Hence, by the principle of conservation of energy, we must have

$$m_0c^2(1-\kappa)^{1/2}/(1-\beta^2)^{1/2} = m_0c^2(1-a/r_1)^{1/2}. \tag{3}$$

Note that the absorption lines are always formed near the star's surface before the motion of the falling matter is damped because this is where the density, and hence the optical thickness, of the falling matter is greatest. Let $\alpha = (1-a/r_1)^{1/2} \leq 1$; from equations (2) and (3), we obtain the Doppler redshift as

$$z_d = [\alpha + (\alpha^2 - 1 + \kappa)^{1/2}]/(1-\kappa)^{1/2} - 1. \tag{4}$$

The maximum Doppler redshift $z_D$ corresponds to $\alpha = 1$, where we have

$$z_D = (1 + \kappa^{1/2})/(1-\kappa)^{1/2} - 1, \tag{5}$$

and from equations (1) and (5), the relationship between $z_D$ and $z_G$ will be

$$z_D = z_G + \sqrt{2z_G + z_G^2}, \tag{6}$$

$$z_G = \frac{z_D^2}{2(1+z_D)}; \tag{7}$$

$$z_D/z_G = 1 + \sqrt{2/z_G + 1} > 2. \tag{8}$$

Clearly, $z_D$ will be much greater than $z_G$ if GRB redshifts are generated by gravity.

It should be emphasized here that gravity-generated redshift can contain both absorption and emission feature, but Doppler redshift will only observe the absorption feature, the Doppler emission lines cannot be seen due to its faintness.

The emission and absorption features of 150 GRBs are recorded in Table 1 of Ref. [24] and the number-redshift relations for the two kinds of GRBs are shown in Fig. 1, where GRBs that have absorption feature only are merged into $N_{abs}(z)$, others into $N_{em}(z)$. Five evidences can be discovered from Fig. 1, which support the view that GRB redshifts are gravity-generated and not cosmological in origin.

**1st evidence** It is well known that the detection of absorption lines is more difficult than that of emission lines for quasars. However, a quasar emission redshift should be determined if its absorption redshift is detected. At present, however, only the absorption redshifts have been detected for over half of the known GRBs with the corresponding emission redshifts mysteriously missing. This enigma can be explained if GRB redshifts are caused by the gravity of neutron stars, where the thickness of the returned matter of most GRB is thick enough to ensure the absorption lines can be clearly seen and making the lower emission redshift cannot be identified. the emission lines can be seen only at GRBs which the Doppler absorption lines are relatively faint.

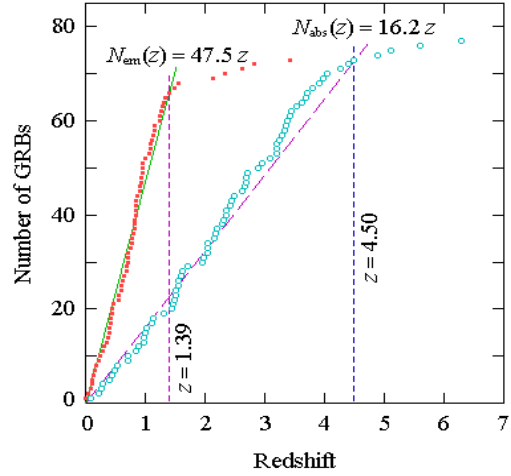

**Figure 1.** Number–redshift relations for GRBs with absorption redshifts (for 77 GRBs) and emission redshifts (for 73 GRBs including 19 GRBs that have redshifts with both emission and absorption features).

**2nd evidence** There are more QSOs with emission redshifts than those with absorption redshifts because the foreground absorbers do not always exist. On the contrary, of the 150 GRBs observed, there are more GRBs with absorption features only than those with emission features. This fact is not in agreement with GRBs being cosmological objects but consistent with the view that GRB redshifts are generated by the gravity of neutron stars.

**3rd evidence** The mean redshift ($\bar{z}_{abs} = 2.3751$) of 77 GRBs with absorption features is much higher than that ($\bar{z}_{em} = 0.86886$) of 73 GRBs with emission features. This fact is completely consistent with the claim that GRB redshifts are the results of gravity because equation (8) implies that $z_D > 2z_G$ and $\bar{z}_D > 2\bar{z}_G$. However, owing to the origin of the absorption lines in the foreground absorbers, cosmological redshifts will give rise to smaller redshifts so that we have $z_{abs} \leq z_{em}$ and $\bar{z}_{abs} \leq \bar{z}_{em}$. Therefore the inequality $\bar{z}_{abs} > 2\bar{z}_{em}$ is irrefutable evidence that GRB is not cosmological events.

**4th evidence** Absorption redshifts consistently underestimate the real distances of cosmological objects because the inequality $z_{abs} \leq z_{em}$, which holds for every cosmological object, inevitably increases the slope of the number–reshift relation

$$\frac{dz_{abs}}{dz_{em}} \leq 1, \quad \frac{dN(z_{em})}{dz_{em}} = \frac{dN(z_{em})}{dz_{abs}} \frac{dz_{abs}}{dz_{em}} \leq \frac{dN(z_{abs})}{dz_{abs}},$$

where $N(z_{em}) = N(z_{abs})$ is the cumulative number for the GRBs. Therefore, if GRB redshifts are indeed cosmological, then the slope of the number–redshift relation that is obtained from the absorption features is definitely not less than that obtained from the emission features. This prediction completely contradicts the observed results of Fig. 1 which shows that the slope of the number–emission redshift relation is almost three times as great as that of the number–absorption redshift relation.

**5th evidence** It is known that the total number of cosmological objects in space is proportional to the cube of the redshift in the lower redshift region and that the distribution of objects in the high redshift region (e.g. when $z > 0.4$) is dependent on the spatial curvature and other evolutionary effects (see also [24]). From Fig. 1, however, we see that the linear relation of the emission redshift distribution has extended to $z_{em} \leq 1.39$, and that of the absorption redshift has even extended to $z_{abs} \leq 4.50$. These results are completely disagreement with the distribution laws of cosmological objects.

Moreover, from equations (1) and (5) we have that

$$\kappa = 1 - \frac{1}{(1+z_G)^2}, \quad \kappa = \frac{[(1+z_D)^2 - 1]^2}{[(1+z_D)^2 + 1]^2}, \qquad (9)$$

where $\kappa = 2GM/r_0 c^2$. A unique value of $\kappa$ is therefore obtained for each fixed $z_D$, or $z_G$, and so we have two kinds of count distribution functions $\varphi_D(\kappa)$ and $\varphi_G(\kappa)$ for the variable $\kappa$ corresponding to the two kinds of GRBs with Doppler redshifts and gravitational redshifts respectively. Due to the identification of the two kinds redshifts being stochastic, this implies that some decisive evidence must exist which will finally confirm the origin of GRB redshift, that is the

**6th evidence** If GRB redshifts are generated by the gravity of neutron stars, then we must have $\varphi_D(\kappa) \equiv \varphi_G(\kappa) \equiv \varphi(\kappa)$ because the distribution law of neutron star for either gravitational redshifts or Doppler redshifts is the same.

Note that if these redshifts are indeed the results of gravity, then many of the existing GRB redshifts would in fact be mistakes because they are provided by the background galaxy and not by GRBs. Only those redshifts obtaining from afterglows are reliable. There are 114 observations with such redshifts in Table 1 of [24], 41 of them possess emission features $z_G$ and 73 of them possess absorption features $z_D$; these redshifts are marked by an asterisk (*) in [24]. The number–redshift relation is shown in Fig. 2a, here the GRB having both the same emission and absorption redshift features simultaneously is merged into $z_G$.

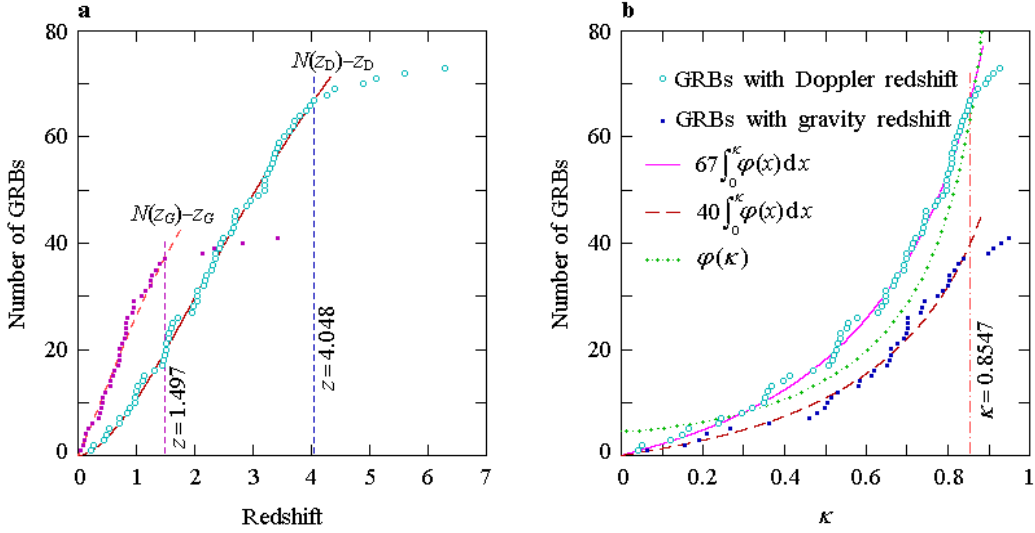

**Figure 2.** Number–redshift relations and number–$\kappa$ relations for GRBs with Doppler redshifts and gravitational redshifts respectively. Fig. 2a shows the $N(z_D)$–$z_D$ relation (solid) and the $N(z_G)$–$z_G$ relation (dash) of GRBs. Fig. 2b shows the number–$\kappa$ relations for GRBs with Doppler redshifts and gravitational redshifts, and the distribution function $\varphi(\kappa)$ (dot-cross). The function $\varphi(\kappa)$ is simultaneously suitable for both kinds of GRB redshifts.

In fact, the number–$\kappa$ relations for the two kinds of GRBs (namely, those with Doppler redshifts and those with gravitational redshifts) are both derivable from equation (9). We find from Fig. 2b that indeed we have $\varphi_D(\kappa) \equiv \varphi_G(\kappa) \equiv \varphi(\kappa)$ for the two kinds of GRBs and

$$\varphi(\kappa) = 0.7161(\kappa^{1.5} + 0.4\kappa^{-0.02})/(1-\kappa)^{0.85}, \ (\kappa \leq 0.8547) \qquad (10)$$

with

$$\int_0^{0.8547} \varphi(\kappa) d\kappa = 1.$$

We can safely conclude, therefore, that GRB redshifts are generated by gravity and are unrelated to its distances.

The above conclusion can be further verified as follows.

**7th evidence** The distribution function $\varphi(\kappa)$ should give the correct distribution laws of both the relations $N(z_D) - z_D$ and $N(z_G) - z_G$ for the two GRB redshifts, and the mean values of $\bar{z}_D$, $\bar{z}_G$ and $\bar{z}_D / \bar{z}_G$.

Using the distribution function $\varphi(\kappa)$ given by equation (10), we deduce that the number−redshift relations for the two kinds of GRBs are

$$N(z_G) = 40 \int_0^{\kappa(z_G)} \varphi(\kappa) d\kappa, \quad N(z_D) = 67 \int_0^{\kappa(z_D)} \varphi(\kappa) d\kappa,$$

where

$$\kappa(z_G) = 1 - \frac{1}{(1+z_G)^2}, \quad \kappa(z_D) = \frac{[(1+z_D)^2 - 1]^2}{[(1+z_D)^2 + 1]^2}.$$

The results shown in Fig. 2a are consistent with these relations for both $N(z_D)-z_D$ and $N(z_G)-z_G$ and show a good agreement between the theory and the observed data. Furthermore, the mean redshift $\bar{z}_D$, $\bar{z}_G$ and $\bar{z}_D / \bar{z}_G$ for $\varphi(\kappa)$ are

$$\bar{z}_G = \int_0^{0.8396} z_G(\kappa) \varphi(\kappa) d\kappa \left( \int_0^{0.8396} \varphi(\kappa) d\kappa \right)^{-1} = 0.72085 \quad \text{(with } z_G \leq 1.497\text{)},$$

$$\bar{z}_D = \int_0^{0.8547} z_D(\kappa) \varphi(\kappa) d\kappa \left( \int_0^{0.8547} \varphi(\kappa) d\kappa \right)^{-1} = 2.1897 \quad \text{(with } z_D \leq 4.048\text{)},$$

$$\bar{z}_D / \bar{z}_G = 3.0376;$$

where $z_G(\kappa) = (1-\kappa)^{-1/2} - 1$ and $z_D(\kappa) = (1+\kappa^{1/2})/(1-\kappa)^{1/2} - 1$, equations (1) and (5) have been used here. The observed data give $\bar{z}_G = 0.71206$ (for $z_G \leq 1.497$ corresponding to $\kappa \leq 0.8396$), $\bar{z}_D = 2.1943$ (for $z_D \leq 4.048$ corresponding to $\kappa \leq 0.8547$) and $\bar{z}_D / \bar{z}_G = 3.0816$. The errors, relative to the theoretical results, are less than 1.5% in all three cases. The results given by $\varphi(\kappa)$, therefore, tally with the observed data not only for the relations $N(z_D)-z_D$ and $N(z_G)-z_G$ but also for the mean values of $\bar{z}_D$, $\bar{z}_G$ and $\bar{z}_D / \bar{z}_G$. In conclusion, the fact that all these results are dependent only on the unique distribution function $\varphi(\kappa)$ is irrefutable evidence that GRB redshifts are arisen from gravity.

Note that the redshift of each GRB above all has a single value, therefore, above results are all statistic. Are there some data in which $z_D$ and $z_G$ exist simultaneously and satisfy relation $z_D = z_G + \sqrt{2z_G + z_G^2}$ in a GRB? The answer is in the affirmative. Here we found two sets of such data as direct evidences for supporting the view of this paper, i.e. the spectrums of GRB 010222 [26] and GRB 030323 [27], in which the original identifications of spectral lines are imperfect. We reconfirm them as follows.

**8th evidence** There exist $z_D$ and $z_G$ which satisfy relation $z_D = z_G + \sqrt{2z_G + z_G^2}$ simultaneously in the spectrum of GRB 010222 [26].

There is a rich and eximious database, i.e. the NIST Atomic Spectra Database which contains tens-thousand spectral lines. Using this spectra data, we reconfirm the spectral line redshifts of GRB 010222 [26] as follows.

**Table 1: the redshifts of absorption lines of GRB 010222**

| $\lambda$ (Å) | ID | $z_D$ or $z_G$ |
|---|---|---|
| 7065.4 | [Cr II] 2852.192 | 1.477183 |
| 6941.7 | [Pb I] 2802.821 | 1.476683 |
| 6924.0 | [Mn I] 2795.641 | 1.476713 |
|  | [Ar II] 4807.363 | 0.440291 |
| 6438.5 | [Fe II] 2599.147 | 1.477159 |
|  | [Pr II] 4469.91 | 0.440409 |
| 6422.4 | [Cr I] 2592.619 | 1.477186 |
|  | [Mn I] 4458.800 | 0.440388 |
| 6405.4 | [V IV] 2585.410 | 1.477518 |
|  | [F II] 4447.775 | 0.440136 |
| 6381.2 | [Pr II] 4430.37 | 0.440331 |
| 6045.0 | [Fe II] 2440.042 | 1.477416 |
|  | [Ne III] 2440.043 | 1.477416 |
|  | [N III] 4196.92 | 0.440342 |
| 6028.2 | [Fe II] 2433.612 | 1.477059 |
| 5900.1 | [V I] 4096.630 | 0.440233 |
| 5879.6 | [Al I] 2373.847 | 1.476824 |
|  | [Zr III] 4081.416 | 0.440577 |
| 5805.4 | [V III] 2343.854 | 1.476861 |
| 5606.7 | [Ne III] 2263.867 | 1.476603 |
|  | [Ba II] 3892.882 | 0.440244 |
| 5574.1 | [Dy I] 3869.91 | 0.440369 |
| 5402.2 | [Zr III] 2180.778 | 1.477189 |
|  | [Fe I] 3750.551 | 0.440375 |
|  | [O II] 3750.550 | 0.440375 |
| 5389.1 | [Ni II] 2175.351 | 1.477347 |
|  | [Ni II] 2175.824 | 1.476740 |
| 5137.2 | [Fe I] 3566.397 | 0.440445 |
|  | [Ne II] 3566.842 | 0.440266 |
| 5108.5 | [Zr III] 2062.137 | 1.477286 |
|  | [Ar II] 3546.608 | 0.440390 |
|  | [Ar II] 3546.856 | 0.440290 |
| 5018.5 | [Zn II] 2026.13 | 1.476889 |
|  | [N IV] 3484.00 | 0.440442 |
| 4478.9 | [Mg IV] 1808.276 | 1.476889 |
| 4137.8 | [Al II] 1670.787 | 1.476562 |
|  | [S III] 2872.800 | 0.440337 |
| 3838.0 | [N V] 1549.335 | 1.477192 |
| 3781.8 | [Si II] 1526.707 | 1.477096 |

Table 1 gives $\bar{z}_D = 1.4770387 \pm 0.000286$ and $\bar{z}_G = 0.4403467 \pm 0.0000977$; and formula (7) claims $\bar{z}_G = 0.4403733$; the relative error is only 0.0060% here!

**9th evidence** There exist $z_D$ and $z_G$ which satisfy relation $z_D = z_G + \sqrt{2z_G + z_G^2}$ simultaneously in the spectrum of GRB 030323 [27].

As above we did, we reconfirm the spectral line redshifts of GRB 030323 [27] by using NIST Atomic Spectra Database as follows.

**Table 2: the redshifts of absorption lines of GRB 030323**

| λ (Å) | ID | $z_D$ or $z_G$ |
|---|---|---|
| 5 415.8 ± 0.31 | [N V] 1 238.821 | 3.371 74 |
|  | [Co II] 2 354.134 | 1.300 55 |
| 5 433.5 ± 0.20 | [N V] 1 242.804 | 3.371 97 |
| 5 465.4 ± 0.15 | [Ti III] 2 375.711 | 1.300 53 |
|  | [Fe II] 2 375.918 | 1.300 33 |
| 5 467.4 ± 0.21 | [V IV] 1 250.918 | 3.370 71 |
|  | [Fe II] 2 377.155 | 1.299 98 |
| 5 481.1 ± 0.29 | [S II] 1 253.811 | 3.371 55 |
|  | [Fe II] 2 382.766 | 1.300 31 |
| 5 505.0 ± 0.33 | [S II] 1 259.519 | 3.370 72 |
|  | [Ne III] 2 393.17 | 1.300 30 |
| 5 509.7 ± 1.00 | [Si II] 1 260.422 | 3.371 31 |
|  | [Ni II] 2 395.252 | 1.300 26 |
| 5 529.2 ± 0.75 | [Ge II] 1 264.710 | 3.371 91 |
|  | [Si II] 1 264.738 | 3.371 81 |
| 5 591.5 ± 0.11 | [Sn I] 2 430.23 | 1.300 81 |
| 5 595.7 ± 0.10 | [V IV] 2 432.623 | 1.300 27 |
| 5 624.4 ± 0.11 | [Ti III] 1 286.365 | 3.372 32 |
|  | [Zr III] 2 445.337 | 1.300 05 |
| 5 634.7 ± 0.16 | [Cl III] 2 449.33 | 1.300 51 |
|  | [Zr III] 2 449.628 | 1.300 23 |
| 5 649.2 ± 0.16 | [Ne III] 2 455.668 | 1.300 47 |
|  | [O III] 2 455.709 | 1.300 44 |
|  | [W I] 2 455.716 | 1.300 43 |
| 5 652.0 ± 0.15 | [W I] 2 457.276 | 1.300 11 |
| 5 692.5 ± 0.69 | [O I] 1 302.168 | 3.371 56 |
|  | [W I] 2 474.898 | 1.300 10 |
| 5 702.4 ± 0.70 | [V IV] 1 304.173 | 3.372 43 |
|  | [Si II] 1 304.370 | 3.371 77 |
|  | [Fe II] 2 479.321 | 1.299 98 |
| 5 709.7 ± 0.32 | [S I] 1 305.884 | 3.372 29 |
|  | [Ne III] 1 306.131 | 3.371 46 |
|  | [W I] 2 482.188 | 1.300 27 |
| 5 723.9 ± 0.29 | [Hg II] 1 309.136 | 3.372 27 |
|  | [Si II] 1 309.276 | 3.371 81 |
| 5 833.7 ± 0.84 | [C II] 1 334.532 | 3.371 35 |
|  | [Fe I] 2 536.369 | 1.300 02 |
|  | [P I] 2 536.374 | 1.300 02 |
| 5 838.9 ± 0.79 | [C II] 1 335.708 | 3.371 39 |
|  | [Cl I] 1 335.726 | 3.371 33 |
|  | [Fe II] 2 537.900 | 1.300 68 |
| 6 092.9 ± 1.08 | [Si IV] 1 393.755 | 3.371 57 |
|  | [Hg II] 1 393.792 | 3.371 46 |
| 6 132.4 ± 1.15 | [Si IV] 1 402.770 | 3.371 64 |
|  | [Fe II] 2 665.457 | 1.300 69 |
|  | [S III] 2 666.233 | 1.300 02 |
|  | [Cl III] 2 666.29 | 1.299 98 |
| 6 278.0 ± 0.42 | [Si III] 1 436.160 | 3.371 38 |
|  | [Li II] 2 729.134 | 1.300 36 |
| 6 629.2 ± 0.14 | [Na II] 2 881.994 | 1.300 21 |
| 6 673.8 ± 0.75 | [Si II] 1 526.707 | 3.371 37 |
| 6 703.7 ± 0.34 | [Si II] 1 533.431 | 3.371 70 |
| | [Hg II] 1 533.590 | 3.371 25 |
| | [Sn I] 2 914.39 | 1.300 21 |
| 6 743.0 ± 0.14 | [V II] 2 931.649 | 1.300 07 |
| | [Cr II] 2 931.704 | 1.300 03 |
| | [Na II] 2 931.793 | 1.299 96 |
| 6 754.2 ± 0.13 | [Cr II] 2 935.991 | 1.300 48 |
| 6 760.6 ± 0.14 | [Bi I] 2 939.16 | 1.300 18 |
| | [K III] 2 939.31 | 1.300 06 |
| 6 766.9 ± 0.99 | [C IV] 1 548.187 | 3.370 85 |
| | [In II] 2 941.91 | 1.300 17 |
| 6 769.9 ± 0.51 | [Mn VI] 1 548.430 | 3.372 11 |
| | [Ti I] 2 942.852 | 1.300 45 |
| 6 778.2 ± 0.84 | [C IV] 1 550.772 | 3.370 86 |
| | [Mo I] 2 946.52 | 1.300 41 |
| 6 781.0 ± 0.53 | [C IV] 1 550.772 | 3.372 66 |
| | [W I] 2 947.846 | 1.300 32 |
| | [Hg II] 2 947.933 | 1.300 26 |
| 6 870.4 ± | [S III] 2 986.860 | 1.300 21 |
| | [K III] 2 987.08 | 1.300 04 |
| 6 876.5 ± | [Mg III] 1 572.713 | 3.372 38 |
| | [Cr I] 2 989.520 | 1.300 20 |
| | [Sc II] 2 989.798 | 1.299 99 |
| 6 891.9 ± | [Cr I] 2 995.973 | 1.300 39 |
| 6 902.4 ± | [Mg IV] 1 578.547 | 3.372 63 |
| | [K II] 1 578.871 | 3.371 73 |
| 7 031.7 ± 0.48 | [Fe II] 1 608.451 | 3.371 72 |
| | [V IV] 3 056.753 | 1.300 38 |
| | [Ne III] 3 057.003 | 1.300 19 |
| 7 043.6 ± 0.17 | [Mg IV] 1 611.214 | 3.371 61 |
| | [K III] 3 062.14 | 1.300 22 |
| 7 082.0 ± 0.15 | [B V] 1 619.969 | 3.371 69 |
| | [V IV] 3 078.370 | 1.300 57 |
| 7 160.1 ± 0.20 | [Ne III] 1 638.177 | 3.370 77 |
| | [Eu I] 3 112.33 | 1.300 56 |
| 7 202.7 ± 0.17 | [Be II] 3 131.329 | 1.300 21 |
| 7 303.5 ± 0.99 | [Al II] 1 670.787 | 3.371 29 |
| | [Zr III] 1 670.927 | 3.370 93 |
| | [F III] 3 175.097 | 1.300 25 |
| 7 907 ± 0.58 | [Mg IV] 1 808.276 | 3.372 67 |
| 8 112 ± 0.45 | [Na III] 1 855.912 | 3.370 90 |
| 8 145 ± 0.23 | [Al III] 1 862.790 | 3.372 47 |
| 8 237 ± 0.34 | [Sc II] 3 581.663 | 1.299 77 |
| 8 802 ± 0.55 | [Fe I] 3 826.966 | 1.299 99 |
| 8 859 ± 0.82 | [Zn II] 2 026.13 | 3.372 38 |
| | [F II] 3 851.077 | 1.300 40 |
| 8 970 ± 0.73 | [Na III] 2 052.143 | 3.371 04 |
| | [K II] 3 899.000 | 1.300 59 |
| | [Mg I] 3 899.163 | 1.300 49 |
| 9 019 ± 0.77 | [Ne III] 2 063.471 | 3.370 79 |
| | [O II] 3 920.382 | 1.300 54 |

Table 2 gives $\bar{z}_D = 3.3716344 \pm 0.0005634$ and $\bar{z}_G = 1.3002702 \pm 0.0002244$ ;

and formula (7) claims $\bar{z}_G = 1.3001909$; the relative error is only 0.0061% here!

We can conclude finally that above results, no matter the indirect statistic result or the direct data, all show unanimously that GRB redshift is generated by gravity and has nothing to do with GRB distance.